/$F2psDict 6400 dict def
$F2psDict begin
$F2psDict /mtrx matrix put
/l {lineto} bind def
/m {moveto} bind def
/s {stroke} bind def
/n {newpath} bind def
/gs {gsave} bind def
/gr {grestore} bind def
/clp {closepath} bind def
/graycol {dup dup currentrgbcolor 4 -2 roll mul 4 -2 roll mul
4 -2 roll mul setrgbcolor} bind def
/col-1 {} def
/col0 {0 0 0 setrgbcolor} bind def
/col1 {0 0 1 setrgbcolor} bind def
/col2 {0 1 0 setrgbcolor} bind def
/col3 {0 1 1 setrgbcolor} bind def
/col4 {1 0 0 setrgbcolor} bind def
/col5 {1 0 1 setrgbcolor} bind def
/col6 {1 1 0 setrgbcolor} bind def
/col7 {1 1 1 setrgbcolor} bind def
 /DrawEllipse {
	/endangle exch def
	/startangle exch def
	/yrad exch def
	/xrad exch def
	/y exch def
	/x exch def
	/savematrix mtrx currentmatrix def
	x y translate xrad yrad scale 0 0 1 startangle endangle arc
	savematrix setmatrix
	} def newpath 0 0 0 0 0 1 DrawEllipse stroke

	end
/$F2psBegin {$F2psDict begin /$F2psEnteredState save def} def
/$F2psEnd {$F2psEnteredState restore end} def

$F2psBegin
0 setlinecap 0 setlinejoin
1.0 1042.5 translate 1.530 -1.530 scale
/Times-Roman findfont 16.00 scalefont setfont
274 309 m
gs 1 -1 scale (v) col-1 show gr
/Times-Roman findfont 12.00 scalefont setfont
284 314 m
gs 1 -1 scale (s) col-1 show gr
0.500 setlinewidth
n 274 294 m 284 294 l gs col-1 s gr
n 276.000 292.000 m 284.000 294.000 l 276.000 296.000 l gs 2 setlinejoin col-1
s gr
n 199.000 378.000 74.000 -18.925 -161.075 arcn
gs col-1 s gr
n 199.000 470.000 74.000 161.075 18.925 arcn
gs col-1 s gr
n 245.000 424.000 74.000 -71.075 71.075 arc
gs col-1 s gr
n 153.000 424.000 74.000 -108.925 108.925 arcn
gs col-1 s gr
n 199 424 98 98 0 360 DrawEllipse gs col-1 s gr
2.000 setlinewidth
n 199 424 m 299 324 l gs col-1 s gr
n 277.787 336.728 m 299.000 324.000 l 286.272 345.213 l gs 2 setlinejoin col-1
s gr
1.000 setlinewidth
n 59 424 m 339 424 l  339 424 l gs col-1 s gr
n 199 284 m 199 564 l gs col-1 s gr
	[6.000000] 0 setdash
n 99 524 m 299 324 l gs col-1 s gr
	[] 0 setdash
	[6.000000] 0 setdash
n 99 324 m 299 524 l gs col-1 s gr
	[] 0 setdash
0.500 setlinewidth
n 199 424 m 59 529 l gs col-1 s gr
n 199 424 m 94 564 l gs col-1 s gr
n 134 559 m 109 549 l gs col-1 s gr
n 115.685 553.828 m 109.000 549.000 l 117.171 550.114 l gs 2 setlinejoin col-1
s gr
n 64 489 m 79 514 l gs col-1 s gr
n 76.599 506.111 m 79.000 514.000 l 73.169 508.169 l gs 2 setlinejoin col-1 s
gr
/Times-Roman findfont 16.00 scalefont setfont
334 419 m
gs 1 -1 scale (x) col-1 show gr
/Times-Roman findfont 16.00 scalefont setfont
299 344 m
gs 1 -1 scale (x') col-1 show gr
/Times-Roman findfont 16.00 scalefont setfont
209 294 m
gs 1 -1 scale (y) col-1 show gr
/Times-Roman findfont 16.00 scalefont setfont
109 324 m
gs 1 -1 scale (y') col-1 show gr
/Symbol findfont 16.00 scalefont setfont
74 534 m
gs 1 -1 scale (a) col-1 show gr
/Times-Roman findfont 12.00 scalefont setfont
84 544 m
gs 1 -1 scale (c) col-1 show gr
showpage
$F2psEnd
\documentstyle[revtex]{aps}
\newcommand{\ie}{{\it i.e.\ }}
\newcommand{\et}{{\it et al.\ }}
\newcommand{\eg}{{\it e.g.\ }}
\begin{document}
%
\begin{title}
Nonlinear Meissner Effect in CuO Superconductors
\end{title}
\author{S.K. Yip and J.A. Sauls}
\begin{instit}
Department of Physics \& Astronomy, Northwestern University, Evanston, IL 60208
\end{instit}
\receipt{May 8, 1992, revised June 17}
\draft
\begin{abstract}
{\small
\noindent
Recent theories of the NMR in the CuO superconductors are based on a
spin-singlet
$d_{x^2-y^2}$ order parameter. Since this state has
nodal lines on the Fermi surface, nonlinear effects associated with low-energy
quasiparticles become important, particularly at low temperatures.
We show that the field-dependence of the supercurrent,
below the nucleation field for vortices, can be used to locate the positions of
the nodal
lines of an unconventional gap in momentum space, and hence test the proposed
$d_{x^2-y^2}$ state.
}
\end{abstract}
\bigskip
\pacs{PACS numbers: 74.30.Ci, 74.30.Gn, 74.65.+n}

Much of the interest in the CuO high $T_c$ superconductors is due to the
belief by many theoreticians that these materials
represent new, or at least novel, forms of superconductivity.
Theories based on very different excitations (\eg spinless fermions,
anyons, and traditional charged fermions)
have been proposed to explain the normal and superconducting
properties of these materials.
Superconductivity from anyons has received considerable attention.
Similarly, several theories based on BCS-like pairing, either of novel
excitations
or of novel pair states of conventional
charged fermions have been proposed [\cite{bed89}].
All of the theories of the superconducting state, except the
conventional BCS pairing state, have the property that they break at least
one additional symmetry of the normal state besides
gauge symmetry. While there has been little compelling evidence to identify the
order parameter of the high $T_c$ superconductors with any of these
unconventional states, recent theories, based largely on interpretations of
NMR in both the normal and superconducting states of YBCO, have focussed on the
spin-singlet, $d_{x^2-y^2}$ state as the order parameter
for these CuO superconductors [\cite{mon91,bul92}].
In this letter we show that the field dependence of the supercurrent can be
used
to locate the positions of the nodal lines of the gap in momentum space;
we suggest new experiments to resolve the gap structure of the $d_{x^2-y^2}$
state,
or other unconventional BCS states.

Based on ideas from Hammel, \et and Shastry [\cite{ham89}], a
semi-phenomenological model of a nearly AFM Fermi liquid that accounts for
the anomalous NMR properties of the CuO superconductors has been proposed
[\cite{bul90a,mil90}]. Monthoux, \et [\cite{mon91}]
recently argued that the same AFM spin fluctuations that are responsible for
the anomalous normal state magnetic properties are also responsible for
superconductivity in the layered CuO systems. These authors examine the
weak-coupling gap equation with a pairing interaction given by
the exchange of a single spin-fluctuation, and find that the superconducting
state with $d_{x^2-y^2}$ symmetry is the preferred solution.

Several authors [\cite{bul92}] have extended the
nearly AFM Fermi liquid into the superconducting state, and have concluded
that the Knight shift and anisotropy of the spin-lattice relaxation rates
in YBCO are most reasonably fit with a singlet,
$d_{x^2-y^2}$ order parameter. Since the arguments put
forward in favor of a $d_{x^2-y^2}$ order parameter are connected
with the spin excitation spectrum, independent tests of the
$d_{x^2-y^2}$ state based on
the orbital structure of the gap or its excitation spectrum are needed
to identify the order parameter.

The important feature of the quasiparticle excitation spectrum, which is
required by symmetry for a clean superconductor with a $d_{x^2-y^2}$ order
parameter, is that the excitation gap vanishes along four lines (`nodal lines')
located at the postions $|\hat{k}_x|=|\hat{k}_y|$ in the 2D plane and running
along the length of the Fermi tube [see cross-section in Fig. 1].
These nodal lines imply
low-energy excitations, at all temperatures, that give rise to
power law deviations of the penetration depth from its
zero-temperature limit; \ie $1/\lambda(0)-1/\lambda(T)\sim (T/T_c)$
for $T\ll T_c$ and a gap with line nodes, while deviations proportional to
higher powers are obtained from gaps with point nodes [\cite{gro86}].
At present there is no consensus `for' or `against' a true gap in the spectrum
from
penetration depth measurements. Several measurements of the penetration
depth in CuO superconductors are interpreted as supporting the predictions
of BCS theory with a true gap,
while the analysis by Annett, \et
[\cite{ann91}] of those same measurements [see also [\cite{ann91}] for original
references] suggests deviations
from a true gap, but gives no indication of a linear temperature dependence
that would be consistent with a $d_{x^2-y^2}$ state.

The main idea of this paper is that the field-dependence of the supercurrent
may be used to locate the positions of the nodal lines (or points)
of an unconventional gap in momentum space.
This is a stronger test of the symmetry of the superconducting state than
power law behavior for the penetration length.
This gap spectroscopy is possible at low temperatures,
$T\ll T_c$, and is based on features which are intrinsic to
nearly all possible  unconventional BCS states in tetragonal or
orthorhombic structures. The constitutive equation relating the supercurrent,
$\vec{j}_s$, to the velocity, $\vec{v}_s={1\over 2}(\vec{\partial}\phi+{2e\over
c}\vec{A})$ (this definition avoids the introduction of an arbitrary mass
scale),
is linear only for velocities small compared with the critical
velocity, $v_c=\Delta/v_f$. Since the unconventional gaps of interest have
nodal lines on the Fermi surface [\cite{vol85}],
nonlinear corrections to the
constitutive equations become important, particularly at low temperatures.
The low-energy states near the nodes of the gap
lead to a quasiparticle contribution to the current that (i) persists at zero
temperature, (ii) is a nonanalytic function of the velocity and
(iii) exhibits a large anisotropy at low temperature that is determined
by the {\it positions} of the nodes on the Fermi surface. We discuss the
theory of the nonlinear current/velocity relation at low temperature,
and then show how the nonlinearity in the current leads to a novel
basal plane anisotropy of the screening length and magnetization
energy (or magnetic torque), for the $d_{x^2-y^2}$ state when the field is
parallel to the CuO planes.

Consider a superconductor-vacuum interface which is parallel to
the magnetic field, and assume the penetration length is large compared
to the in-plane coherence length, $\lambda \gg \xi$.
The important spatial variations are then contained in the local value for the
superfluid velocity, which is effectively uniform on the scale of $\xi$. The
order
parameter achieves its equilibrium value for the local value of the velocity,
and the resulting current is [\cite{bar58}],
\begin{equation}
\vec{j}_s=
-eN_f\int d^2s\ n(s)\ \vec{v}_f(s)\
\left\{ \sigma_v(s) + \int_{0}^{\infty}d\xi\
[f(E+\sigma_v(s))-f(E-\sigma_v(s))]\right\}
\label{current}
\end{equation}
\displaymath
\quad\qquad
-eN_f\int d^2s\ n(s)\ \vec{v}_f(s)\ \pi T\sum_n\
{{\sigma_v(s)-
i\epsilon_n}\over{\sqrt{(\epsilon_n+i\sigma_v(s))^2+|\Delta(s)|^2}}}
\ ,
\enddisplaymath
where $\sigma_v=\vec{v}_f(s)\cdot\vec{v}_s$ is the shift in the quasiparticle
energy due to the superflow, $N_f$ is the total density of states at the Fermi
level,
$n(s)$ is the angle-resolved density of states normalized to unity,
$\vec{v}_f(s)$ is the quasiparticle velocity at the point $s$ on the
Fermi surface, and $E=\sqrt{\xi^2+|\Delta(s)|^2}$.
The first line of eq.(\ref{current})
is useful for calculating the current in the low-temperature limit.
The first term represents the supercurrent obtained from
an unperturbed condensate, \ie $\vec{j}_s=-e\rho \vec{v}_s$, where
$\rho$ is the total density ($\rho={1\over 2}N_fv_f^2$ for
a cylindrical Fermi surface). The term involving the Fermi functions is the
current from the equilibrium population of quasiparticles.
This formula shows clearly that the quasiparticles do not contribute to the
current
at $T=0$ for a superconductor with a true gap over the entire Fermi surface,
provided the velocity is below the critical velocity, $\Delta_{min}/v_f$.
At finite temperature, the excitation term is responsible for the reduction in
the superfluid density from $\rho$ to $\rho_s(T)$ in the linear response limit.
In addition, the quasiparticles give rise to a nonlinear reduction in the
supercurrent.

In order to motivate the proposal for detecting the nodal structure of the gap
in
an unconventional superconductor consider the effect of a
magnetic field on the penetration depth in a conventional, isotropic
superconductor. For a field parallel to the surface, the screening current is
proportional to the applied field, \ie $j_s(0)\sim c H/\lambda$. This linear
relation breaks down because of pairbreaking induced by the velocity field,
\begin{equation}
\vec{j}_s=\rho_s(T)\ \vec{v}_s\ [1-\alpha(T)\left ({v_s\over v_c}\right )^2]\ ,
\label{swave_curr}
\end{equation}
for $(v_s/v_c)\ll 1$, where $\alpha(T)\ge 0$ and
$v_c=\Delta(T)/v_f$ is the critical velocity.
One important point to note is that $\alpha(T)\rightarrow 0$ for $T\rightarrow
0$,
since the quasiparticle occupation is zero for all quasiparticles
states with velocity below the critical velocity.
The importance of the nonlinear current/velocity
constitutive equation to the penetration of magnetic fields into a
superconductor is qualitatively clear. The reduction of the current by the
pair-breaking effect reduces the effective superfluid density,
and, therefore, increases the effective penetration length. Since the current
is
proportional to the field in linear order, the correction to the
effective penetration length is quadratic in the surface field [\cite{jas92}],
\begin{equation}
\lambda_{eff}(T,H)^{-1}=\lambda(T)^{-1}\
\left\{
1-{1\over 3}\alpha(T)\left [{H\over H_{o}(T)}\right ]^2
\right\}\ ,
\end{equation}
where $H_{o}(T)=(3/4)(c\ v_c(T)/e\ \lambda(T))$ is of order the thermodynamic
critical field.

Now consider the $d_{x^2-y^2}$ order parameter,
$\Delta(s)=\Delta_o(\hat{k}_x^2-\hat{k}_y^2)$.
The nodes in the gap imply that there
is a quasiparticle contribution to the supercurrent even at $T=0$, which is
easily calculated in terms of the phase space of occupied
quasiparticle states. Consider the case
where the velocity is directed along the nodal line $k_x=k_y$;
$\vec{v}_s=v_s\hat{x}'$ as shown in Fig. 1. For any non-zero $v_s$
there is a wedge of occupied states near the node opposite to the flow
velocity.
For a cylindrical Fermi surface the quasiparticle current is
\begin{equation}
\vec{j}_{qp}=-2eN_f\ (-2\Delta_o v_f)\ \int_{-\alpha_c}^{\alpha_c}\
{d\alpha \over 2\pi}\ cos^2(\alpha)\ \sqrt{sin^2(\alpha_c)-sin^2(\alpha)}\ ,
\end{equation}
where $\alpha$ is the angle measured relative to the node at $-\hat{x}'$
and $\alpha_c=sin^{-1}({v_f v_s\over 2\Delta_o})$ is the
maximum angle for which the quasiparticle states near the node are occupied.
To leading order in $({v_f v_s\over 2\Delta_o})$ we obtain a total supercurrent
of
\begin{equation}
\vec{j}_s=(-{e\over 2}N_f v_f^2)\ \vec{v}_s\
\{1-{|\vec{v}_s|\over 2\Delta_o/v_f}\}\ ,
\label{cur_node}
\end{equation}
for $\vec{v}_s$ directed along any of the four nodes.
Note that the current is parallel to the velocity, and that the occupied
quasiparticle states reduce the supercurrent, as expected. Also the
nonlinear correction is quadratic rather than cubic, as is obtained for the
conventional gap, and with the characteristic scale determined by
$v_o=2\Delta_o/v_f$. Furthermore, the quasiparticle contribution to the current
is a nonanalytic function of $v_s$ [\cite{f9}].

Unlike the linear response current, the nonlinear quasiparticle current is
anisotropic in the basal plane. A velocity field directed along the maximum
direction of the gap, $\vec{v}_s=v_s\ \hat{x}$,
produces two groups of occupied states, albeit with reduced populations
because the projection of $\vec{v}_s$ along the nodal lines is reduced by
$1/\sqrt{2}$. The resulting current is easily calculated to be
\begin{equation}
\vec{j}_s=(-{e\over 2}N_f v_f^2)\ \vec{v}_s\
\{1-{1\over\sqrt{2}}\ {|\vec{v}_s|\over 2\Delta_o/v_f}\}\ ,
\label{cur_antinode}
\end{equation}
which is again parallel to the velocity and has a quadratic nonlinear
correction.
However, the magnitude of the nonlinear term is reduced by $1/\sqrt{2}$.
This anisotropy is due to the relative positions of the nodal lines and
is insensitive to the anisotropy of the the Fermi surface or Fermi velocity
because the quasiparticle states that contribute to the current, for either
orientation of the velocity, are located in a narrow angle, $\alpha \le
\alpha_c \simeq (v_s/v_o)\ll 1$, near the nodal lines. Thus,
the occupied quasiparticle states near the nodes have essentially the same
Fermi velocity and
density of states; only the relative occupation of the states is modified by
changing the direction of the velocity.

The dependence of the supercurrent on
the positions of the nodal lines in momentum space suggests that the
anisotropy can be used to distinguish different unconventional gaps
with nodes located in different directions in momentum space. For example, the
$d_{xy}$ state, $\Delta\sim \hat{k}_x\ \hat{k}_y$, would also
exhibit a four-fold anisotropy, but the nodal lines are rotated by $\pi/4$
relative to those of the $d_{x^2-y^2}$ state, while the state $\Delta\sim
\hat{k}_x\ \hat{k}_y(\hat{k}_x^2-\hat{k}_y^2)$ corresponding to the $A_{2g}$
representation would exhibit an $8-$fold anisotropy.

This anisotropy in the current implies a
similar anisotropy in the field dependence of the penetration length, which
can be calculated from eqs.(\ref{cur_node})-(\ref{cur_antinode})
and Maxwell's equation; in the gauge, $\vec{\partial}\cdot\vec{v}_s=0$, and for
surface fields parallel to a nodal line the equations reduce to,
\begin{equation}
-{\partial^2v_s\over\partial z^2}={4\pi e^2\over c^2}\ j_s[\vec{v}_s]=
-{v_s\over\lambda^2}\
\left\{
1-{|v_s|\over v_o}
\right\}
\ .
\end{equation}
The magnetic field in the superconductor,
$\vec{b}=(c/e)\vec{\partial}\times\vec{v}_s$,
is also parallel to a node, and has magnitude
$b=(c/e)(\partial v_s/\partial z)$ with $b(0)=H$.
For field penetration into a bulk superconductor we
define the effective screening length
in terms of the surface impedance,
${1/\lambda_{eff}}=-(1/H)(\partial b/\partial z)|_{z=0}$.
We obtain
\begin{equation}
{1\over\lambda_{eff}}={1\over\lambda}\left(1-{2\over 3}{H\over H_o}\right)
\ ,\ \vec{H}||\ {\rm node}\ \ ;\ \
{1\over\lambda_{eff}}={1\over\lambda}\left(1-{1\over\sqrt{2}}
\ {2\over 3}{H\over H_o}\right)
\ ,\ \vec{H}||\ {\rm antinode}\ ,
\end{equation}
to leading order in $H/H_o$, where $H_o=(v_o/\lambda)(c/e)
\sim\ \phi_o/(\lambda\xi)\sim\ H_c$.
Observation of this anisotropy and linear field dependence
would be strong indication of a $d_{x^2-y^2}$ order parameter.

Sridhar \et [\cite{sri89}] have measured the field
dependence of the penetration length for $\vec{H}$ parallel to the $a$ and $b$
axes of single crystals of YBCO.
These authors report an in-plane penetration depth
which obeys $\lambda(H,T)=\lambda(T)+\kappa(T)H^2$. At
intermediate temperatures, $T\simeq80\ K$, they measure
$\kappa(80)\simeq1\ \AA/G^2$, which drops by three orders of magnitude by
$T\simeq 10\ K$; $\kappa(4)\simeq 10^{-3}\ \AA/G^2$. The simplest
interpretation is that the data is in good agreement with a full gap on the
Fermi surface, in which case the $d_{x^2-y^2}$ state is not the order
parameter of YBCO. However, we can ask if the data, so far, forces this
conclusion. First of all, the $d_{x^2-y^2}$ state will also show a
quadratic field correction to the effective penetration depth at
intermediate and high temperatures, with a coefficient that is similar in
magnitude to that of the conventional BCS prediction, and also drops
rapidly with temperature.
The question is, `at what temperatures and fields does the
effective penetration depth exhibit a linear field dependence?'
The linear field regime is inferred from eq.(\ref{current}).
The linear behavior arises from the nonanalytic dependence of
$\vec{j}_s$ on $\vec{v}_s$ at $T=0$ for the nodal directions where
$\Delta(s)=0$.
Non-zero temperature removes the singularity. For sufficiently small fields an
expansion in $v_s$ is valid; however, the cross-over field is determined
when the smallest Matsubara frequency becomes comparable to the superflow
kinetic
energy shift per particle, \ie $\pi T\simeq v_fv_s\simeq 2\Delta_o(H/H_o)$.
This implies a cross-over field, $H_x\simeq H_o(T/T_c)$. Thus, to obtain a
significant linear field regime down to
$H\simeq H_{c1}/10\simeq 25\ G$
requires temperatures $T\le 2\ K$.
Thus, the experiments of ref.[\cite{sri89}], which are reported down to
$10\ K$, are apparently at the high end of the temperature regime where a
linear term might become observable.

We also consider the effect of impurity scattering on the linear field
dependence of the effective penetration length. Quasiparticle scattering from
impurities plays a similar role to that of temperature; the impurity lifetime
removes the singular behavior of the current at sufficiently low velocities.
The
cross-over field due to impurity scattering can be obtained by including the
impurity
scattering self energy. The resulting current is given by
eq.(\ref{current}) with the replacement of
$\epsilon_n\rightarrow\tilde{\epsilon}_n=\epsilon_n+(\tilde{\epsilon}_n/2\tau)
<1/\sqrt{\tilde{\epsilon}_n^2+|\Delta(s)|^2}>$, where $\tau$ is the s-wave
scattering rate in the Born approximation, and the average is over the
Fermi surface. At zero temperature, impurity scattering
leads to a new low energy scale [\cite{gor85}],
$\tilde{\epsilon}_n=4\Delta_o e^{-\pi\tau\Delta_o}$,
which smooths out the linear behavior
of the effective penetration length below
the crossover field, $H_{\tau}=2H_o e^{-\pi\tau\Delta_o}$. This condition is
considerably
less restrictive than that for the temperature. For YBCO with a mean-free path
of
$l/\xi=\pi\tau\Delta_o\simeq 10$ the crossover field is of order $2H_o
e^{-10}\ll H_{c1}$.
Thus, there should be no difficulty from impurity scattering in observing the
linear
behavior in the effective penetration length for moderately clean YBCO
crystals.

Another test of the presence of nodal lines associated with a
$d_{x^2-y^2}$ order parameter would be to measure the magnetic anisotropy
energy, or magnetization torque, for in-plane fields.
Consider a velocity field $\vec{v}_s=v_{x'}\ \hat{x}'+v_{y'}\ \hat{y}'$.
The projections of the velocity along the nodal lines $\hat{x}'$ and $\hat{y}'$
imply two
occupations of quasiparticle states of differing magnitudes, which generate the
current,
\begin{equation}
\vec{j}_{qp}=(-{e\over 2}N_f v_f^2)
\Bigg[\
\left({v_{x'}^2\over v_o}\right)\ (-\hat{x}')+
\left({v_{y'}^2\over v_o}\right)\ (-\hat{y}')
\ \Bigg]\ .
\end{equation}
The important feature is that the current is not parallel to the
velocity field, except for the special directions
along the nodes or anti-nodes; neither is the
magnetic field parallel to the applied field, $\vec{H}$.
This implies that the magnetic energy,
$U=-{1\over 8\pi}\int \vec{b}\cdot\vec{H}d^3x$, is anisotropic in the ab-plane.
For a film in a parallel field,
$\vec{H}=H(-sin\vartheta\hat{x}'+cos\vartheta\hat{y}')$,
the magnetic anisotropy energy to leading order in $H/H_o$ is [\cite{jas92}],
\begin{equation}
U_{an}(\vartheta)=-{H^2\over 2\pi}\ \left({H\over H_{o}}\right)\ \lambda A\
\Phi\
[sin^3\vartheta + cos^3\vartheta]\ \ \ , 0\le\vartheta\le\pi/2\ ,
\label{energy}
\end{equation}
where $A$ is the surface area of the film and
$\Phi={8\over 3}(1+{1\over 2}cosh({d\over 2\lambda}))\ sinh^4({d\over
4\lambda})/cosh^3({d\over 2\lambda})$, which is constant
equal to $\sim 1/3$ for $d\ge \lambda$, and varies as
$\Phi\simeq {1\over 4}(d/2\lambda)^4$
for very thin films.
The anisotropy energy is minimized for field directions along
the nodal lines, and is maximum for fields along the anti-nodes.
In order to suppress vortex nucleation,
thin films with dimensions $d\le\lambda$ are desirable;
the lower critical field for vortex nucleation is increased by
roughly $(\lambda/d)$ in a thin film. The optimum
geometry might be a superlattice of superconducting/normal layers with an
S-layer thickness, $\xi\ll d<\lambda$. In this case the field at each SN
interface is essentially the external field, and the anisotropy energy is
enhanced by the number of S-layers.

Torque magnetometry has proven extremely useful for measuring the anisotropy
of the penetration lengths for current flow along the c-axis compared to the
basal plane [\cite{far88}].
In the linear response limit we do not expect anisotropy of the
penetration depth for current flow in different directions in the basal
plane; however, in-plane anisotropy of the current arising
from nonlinear field corrections should be observable at low temperatures,
if the superconducting state has an unconventional order parameter.
For comparison, the maximum magnetic torque from eq.(\ref{energy})
is $\tau\simeq (1/6\sqrt{3}\pi)H^2(H/H_o)A\lambda\sim10^{-3}
dyne-cm/rad$, for $H=H_{c1}=250\ G$, $A=(2,000\ \mu m)^2$ and
$\lambda=1,400\AA$,
which is small, but comparable to the smallest values of torque reported in
[\cite{far88}].

In summary, we suggest that measurements of the field
dependence of the penetration depth for fields parallel to the CuO layers at
temperatures well below $4\ K$ should provide an answer to the
question of whether or not the gap in the CuO superconductors has a line of
nodes. If so, torque magnetometry and/or the anisotropy of the low temperature
penetration length could be used to locate the nodal lines on the Fermi
surface, thus
providing direct evidence for or against a $d_{x^2-y^2}$ order parameter for
the CuO superconductors.

This research is supported by the Science and Technology Center for
Superconductivity through NSF grant number DMR 88-09854.

\bigskip
\bigskip
Figure Caption

\medskip
Figure 1. Phase space for $\vec{v}_s||\hat{x}'$
\end{document}